\documentclass[aps,pre,superscriptaddress,reprint,showpacs]{revtex4-1}
\usepackage{color}
\usepackage{hyperref}
\hypersetup{colorlinks=true,linkcolor=blue,citecolor=blue,filecolor=blue,urlcolor=blue}
\usepackage{graphicx}
\usepackage{amsmath}
\usepackage{mathptmx}
\usepackage{aas_macros}
\usepackage{enumitem}
\usepackage{multirow,bigstrut,bigdelim}
\begin{document}
\title{Perspectives on astronomy: probing Norwegian  pre-service teachers and middle school students}
\author{Christine Lindstr\o m}
\email{christine.lindstrom@hioa.no}
\address{Faculty of Education and International Studies, Oslo and Akershus University College, PB 4 St. Olavs plass, N-0130 Oslo, Norway}
\author{Vinesh~Rajpaul}
\address{Department of Physics, University of Oxford, Oxford OX1 3RH, UK}
\author{Morten Brendehaug}
\address{Faculty of Education and International Studies, Oslo and Akershus University College, PB 4 St. Olavs plass, N-0130 Oslo, Norway}
\author{Megan C.\ Engel}
\address{Department of Physics, University of Oxford, Oxford OX1 3RH, UK}
\date{September 25, 2015}
\begin{abstract}
We report on ongoing work to gain insight into the astronomy knowledge and perspectives of pre-service teachers and middle school students in Norway. We carefully adapted and translated into Norwegian an existing instrument, the Introductory Astronomy Questionnaire (IAQ); we administered this adapted IAQ to (i) pre-service teachers at the largest teacher education institution in Norway, and (ii) students drawn from eight middle schools in Oslo, in both cases  before and after astronomy instruction. Amongst our preliminary findings---based on an analysis of both free-response writing and multiple-choice responses---was that when prompted to provide responses to hypothetical students, the pre-service teachers exhibited a marked drop in \textcolor{black}{pedagogical} responses pre- to post-instruction, with corresponding shifts towards \textcolor{black}{more authoritative} responses. We also identified potentially serious issues relating to middle school students' conceptions of size and distances in the universe, with significant stratification along gender lines.
\end{abstract}
\maketitle
\section{Introduction} \label{sec:intro}
The Introductory Astronomy Questionnaire, hereafter IAQ, is an instrument developed by \citet{Rajpaul2014}, originally administered as pre- and post-tests to students enrolled in the introductory astronomy course at the University of Cape Town (UCT).  The motivation for the development of the original IAQ was threefold: to optimize the teaching of an astronomy course, and to maximize meaningful engagement with its cohort of students; to investigate whether a simple questionnaire could be used to identify potential candidates for teaching intervention programmes; and to study the ways in which students' views changed after having taken an astronomy course, including \emph{non-content} issues regarding worldview and scientific thinking \cite{Wallace:2013}.

The IAQ comprises a small number of questions probing three main areas of interest: student motivation and views on astronomy and physics; astronomy content; and worldview. While a multiple-choice question format may have proven more convenient for analysis, the IAQ was modelled on the Physics Measurement Questionnaire \citep{Volkwyn:2005,Volkwyn:2008} in which free-response writing formed the basis for the substantive analysis. A number of questions were framed as debates between students, prompting respondents to take a side in and contribute to the debate; the instrument also contained a ranking task \cite{2006AEdRv...5a...1H}, an explaining task (with particular attention paid to perceived audience \citep{Allie:2008,Allie:2010}), and a free-association task. This facilitated the expression of views not suggested by the literature, and mitigated the problem of students misinterpreting questions \cite{Nwosu:2012}.

We adapted the IAQ and translated it into Norwegian, with the following targets in mind: (i) pre-service science teachers, before and after astronomy instruction; and (ii) middle school students, before and after astronomy instruction. (Pre-service teachers are students training at a tertiary institution to become teachers.) Our adapted instrument will hereafter be referred to as the Norwegian IAQ (NIAQ).

We aimed to gain insight into the same areas of interest as before, for both the pre-service teachers and the middle school students. Taking a longer-term view, improving the future teachers' understanding of astronomy content and concepts---and indeed general enthusiasm for astronomy and science---could lead to improving the knowledge and perceptions of future generations of young students in Norway.


\section{Context and sample} \label{sec:course}
The Norwegian education system comprises elementary school (years 1--7), middle school (years 8--10) and high school (years 11--13). More than 97\% of elementary and middle school students are enrolled in public schools \cite{encyclo2015}. Science is compulsory in years 1--11, whereas in years 12 and 13, physics, chemistry and biology---which are separate subjects in these years---are non-compulsory. Science contains one standardized astronomy module in middle school, generally covered in year 9. Norwegian tertiary education is offered in universities and university colleges (the distinction between which is today largely historical). The higher education system is in accordance with the Bologna Process, with three-year bachelor's degrees and two-year master's degrees, although some degree programs deviate from this norm by taking four years to complete.

To qualify as a middle school science teacher in Norway, the most common route is to complete a four-year Bachelor of Teaching degree, which includes the equivalent of one year of science (Science I and II). Middle school teachers are also qualified to teach in upper elementary school (years 5--7). Pre-service elementary school science teachers are required to take Science I, whereas Science II is optional. Consequently, \textcolor{black}{the cohort enrolled in Science II comprises a mixture} of future middle and elementary school teachers. 

The pre-service teachers in this study all attended the largest teacher education institution in Norway, viz.\ Oslo and Akershus University College (HiOA). They had passed Science I and were enrolled in Science II. Both courses comprised the subjects Physics, Chemistry, Biology, Technology \& Design, Meteorology \& Geology, and Science Education. Although these subjects are nationally determined, the detailed curriculum of each subject varies between institutions. Physics I and II at HiOA constitute approximately 20\% of Science I and II. Three consecutive 45-minute lessons are delivered in each three-hour class. Physics I comprises seven such classes (plus one class used for group presentations) and covers thermodynamics, gravity \& buoyancy, sound, light, kinematics, forces and energy; Physics II comprises ten classes and covers electromagnetism (electricity, magnetism and induction), atomic and nuclear physics (atomic physics, nuclear physics and radiation physics) and astronomy (the Sun--Moon--Earth system, the Solar System, and the universe, plus one outdoor observing night).

As for the middle school students, a total of eight non-randomly selected schools in and around Oslo agreed to participate in the study during fall 2014. None of the year 8 students and all of the year 10 students in these schools had studied the \textcolor{black}{standardized} astronomy module, so the year 8 students were considered the pre-test sample, and the year 10 students \textcolor{black}{(from the same schools),} the post-test sample. 

\section{Development of the new questionnaire} \label{sec:development}
The original instrument had eight questions (in some cases comprising sub-questions), some of which were not suitable for the NIAQ because they were specific to a South African undergraduate audience (e.g.\ one question focused on radio astronomy with the Square Kilometer Array telescope). We therefore supplemented the most interesting and pertinent questions from the original IAQ with a few new questions.

A full list of questions will be presented elsewhere; the main themes covered by the NIAQ questions were as follows: (1) opinions on how much there is left to discover in both physics and astronomy, and how interesting and important to society they are; (2) astronomy vs.\ astrology; (3) the possibility of life elsewhere in the universe; (4) the Big Bang as a theory; (5a) size rankings and (5b) simple explanations of five different astronomical objects, viz.\ galaxy, planet, star, universe, solar system; (6) the way in which astronomers learn things about the universe; (7) ranking in terms of distance from the Earth's surface of ten different items, e.g.\ the Sun, the Moon, the center of the Milky Way; and (8) the motion of the Earth around the Sun.

Questions 1, 2, 4, 5 have direct counterparts in the original IAQ, while questions 3 and 7 were piloted prior to the present work (in a second-iteration of the IAQ) with a sample of about a hundred undergraduate astronomy students at UCT. With the exceptions of 1, 5a, and 7, all questions solicited free-response writing in response to debates between, or questions from, hypothetical students.

\textcolor{black}{All questions were translated from English into Norwegian by two native Norwegian speakers: the first author (C.L.) and, independently, a high school chemistry and mathematics teacher (with experience studying in England). The two translated versions were compared and discussed, and where there were differences, the most appropriate wording for the Norwegian context was jointly agreed upon. For example, `A group of year 9 students is having an argument' was translated to `En gruppe 10.-klassinger har en diskusjon', which translates directly as `A group of 10th graders have a discussion'. The syntax is more appropriate in Norwegian, and the meaning is preserved as year 9 students are in their last year of compulsory high school education in South Africa, whereas the equivalent in Norway is year 10. The consensus version was given to three Masters level students in Science Teacher Education to check that the questions were clear; no changes were required. }

\section{Methodology}\label{sec:methods}

The NIAQ was administered to two separate cohorts of pre-service teachers enrolled in Physics II in 2013--2014 (one cohort each semester). The cohorts numbered 41 pre-service teachers in total (24 females and 17 males). Almost \textcolor{black}{all were} ethnically Norwegian, so \textcolor{black}{they} were not asked for demographic information, as this would have uniquely identified any \textcolor{black}{non-ethnic Norwegians}. 

The pre-test was administered during the atomic \& nuclear physics module to both cohorts, whereas the post-test was given to all \textcolor{black}{pre-service teachers} during spring 2014, which was after the examination for the fall cohort and after the main astronomy module, but before the observation night and examination for the spring cohort.  \textcolor{black}{Participants} were given 45 minutes to complete the questionnaire. The pre- and post-tests were completed by 40 and 38 \textcolor{black}{pre-service teachers} respectively. No differences were found between the cohorts during the analysis, so they are not separated in the results.

For the middle school administration, students completed questions 1, 5 and 7. The students were given approximately 20 minutes (year 8) and 25 minutes (year 10) to complete the questionnaire, which allowed nearly all students to complete all questions. A total of 535 year 8 students and 387 year 10 students completed the NIAQ. \textcolor{black}{Of the students that elected to specify their gender, 452 were male and 452 were female.}

All pre-service teacher responses, except in the case of multiple choice questions and ranking tasks, were translated from Norwegian into English. The same person involved in the initial translation of the instrument independently translated a small sample of student responses that was also translated by C.L. V.R.\ then compared the two translations, and found a very high level ($>90\%$) of agreement. Subsequently, the remaining responses were translated by C.L.\ only.  However, frequently text was translated indicating where two different words or expressions could be used, or denoting with a question mark where it was unclear how to translate. All translated responses were read by two native English speakers, and any ambiguities or disagreements were discussed and addressed. Analyzing translated responses limited the grain size of the analysis: we had to consider (our interpretation of) the intended meaning without getting unnecessarily caught up in the details of the specific phrasing. However, in the end, this was considered a benefit, because it was ultimately the meaning conveyed by language that was the focus of the analysis.

Each student was assigned a unique numerical code, which was appended to each page of their NIAQ submissions. Submissions were separated by question, and the analysis was carried out on a question-by-question basis (rather than a student-by-student basis). 

The analysis of free response writing was carried out using an approach suggested by grounded theory \citep{Corbin:1990,Strauss:1997}. A number of `main points' were extracted from each student's written response; thence a list of fine-grained categories, covering one or more main points, was drawn up. As most \textcolor{black}{responses} to a particular question contained more than one main point, they ended up being decomposed into more than one fine-grained category. Following an iterative process in which C.L., V.R.\ and M.C.E.\ compared their results and refined their fine-grained categorization, an average agreement between category assignments of above $90\%$ was obtained.  Again using an iterative process, the fine-grained categories were used to construct broader categories of ideas or responses. These broad emergent categories formed the basis of the results for these questions. As an example of this process, consider the `Big Bang as a theory' question. Some fine-grained categories classified students' definitions of `theory' as colloquial, dogmatic, equivocal, or unclear; on the other hand, others grouped together responses providing specific scientific evidence (e.g.\ expansion of the universe, cosmic microwave background radiation), or those providing a scientific definition of `theory'. These fine-grained categories were subsumed into a broader category that stratified responses as non-scientific vs.\ scientifically-compatible. Another broad category that emerged from the responses was that of `pedagogical' vs.\ `authoritative'; see Section \ref{sec:results} for more details.

The analysis for the ranking and explaining tasks entailed capturing all \textcolor{black}{responses} and identifying incorrect ranking sequences, for the former question, and, for the latter question, capturing responses and assigning scores for the given explanations using an `incorrect/partially correct/correct' metric.  The analysis for the remaining questions entailed simply capturing multiple-choice responses.
\section{Results} \label{sec:results}
At the time of writing, analysis of the NIAQ data is ongoing; full results and associated analyses will be presented in one or more separate papers. Below we simply sketch a few of our (preliminary) findings.

Pre- to post-course, significant gains were seen in the pre-service teachers' ability to provide brief explanations of astronomical objects (stars, planets, and so on), as well as to provide detailed explanations of more complex concepts (e.g.\ the fact that the Earth remains in a stable orbit around the sun rather than spirals into it). \textcolor{black}{They} also appeared to develop a more nuanced view of the nature of astronomy, though further analysis is required before definitive conclusions can be drawn. \textcolor{black}{On the other hand, a striking lack of improvement was noted for the middle school students: on all aspects of questions 5a and 5b, the scores of the year 10 students were either statistically identical to or even slightly lower than those of their year 8 counterparts. This may be attributable to students forgetting what they learnt in their astronomy module or not learning much in the first place, or to the possibility that the NIAQ questions targeted different knowledge to that covered in the astronomy module. Detailed analysis of module content is necessary before further conclusions can be drawn.}

Concerning student views on physics and astronomy, increases were observed in the pre-service teachers' opinions on (i) how interesting, and (ii) how important to society physics and astronomy are, and (iii) how much there is left to discover in these fields; these gains were not, however, statistically significant. Nevertheless, all of the aforesaid opinions started from a high baseline, the average in all cases being between $3.5$ and $4.5$ out of $5$. \textcolor{black}{Our finding for the middle school students was the same: opinions started from similarly high baselines, and did not change significantly pre- to post-instruction.}

Given the weak knowledge of astronomical distance scales presented elsewhere in the literature \citep[see, e.g.,][]{trumper2001,Trumper:2001b}, we were not surprised to find that sizeable numbers of middle school students held incorrect views about sizes and distances in the universe. For example, both before and after instruction, a large number thought that the radius of the Earth is smaller than the height of the Earth's atmosphere ($>55\%$), that the Pole star is contained within the Solar System ($>60\%$), and that planets are larger than stars ($>40\%$). \textcolor{black}{Moreover, the students' performance was significantly stratified along gender lines, with female students scoring on average $15$--$20\%$ lower (depending on which metric was used to grade the ranking tasks) than the male students. Though these and other incorrect views were present also in our sample of pre-service teachers, they fared better pre-instruction (prevalence of $30\%$, $40\%$ and $15\%$, respectively, for the aforesaid incorrect ideas), and showed more significant improvements post-instruction (prevalence of $20\%$, $25\%$ and $8\%$, respectively).}

An unexpected finding was that on questions where the pre-service teachers were asked to `write a detailed account' of what they would say to three year 10 students taking differing positions on various topics (the Big Bang as a theory; astronomy vs.\ astrology), the pre-service teachers exhibited a marked drop in \textcolor{black}{pedagogical} responses pre- to post-instruction. Any indication in a response that the pre-service teacher considered the hypothetical students was categorized as \emph{pedagogical}, whereas all other responses, focusing only on a factual answer to the scientific question, were categorized as \emph{authoritative}. Pre-instruction,  more than a third responded in a pedagogical manner, displaying a clear focus on helping students learn and a willingness to admit deficiencies in their own knowledge; post-instruction, however, fewer than $10\%$ of responses exhibited such characteristics, with the remaining majority giving \textcolor{black}{authoritative} responses (regardless of scientific compatibility), not even making any reference to the hypothetical students. 

A pre-course question that was previously found \citep{Rajpaul2014} to be a good predictor of post-course success, viz.\ question 5b, where students were asked to provide simple explanations of a few astronomical objects, was in our study also found to be a statistically significant predictor of post-course success for the pre-service teachers, as measured both by post-course performance on the NIAQ and post-course Force Concept Inventory \citep{hestenes1992} scores. (\textcolor{black}{To help identify such predictors, we developed numeric encoding schemes for responses to all questions; we then examined systematically the correlation coefficients between responses for all possible pairs of questions.}) Question 8, concerning the reason the Earth orbits rather than spirals into the sun, also turned out (unsurprisingly, perhaps) to be a strong predictor of post-course FCI scores. Also of note was the finding that the pre-service teachers' scores on the different components of the aforementioned explaining task (i.e.\ scores for all individual objects) were remarkably similar to those of the diverse sample of South African students studied by \citet{Rajpaul2014}. This suggests the possibility of consistency in students' understanding of certain astronomical objects (stars, planets, galaxies, and so on) across national, linguistic, and cultural boundaries.

Finally, we found that both pre- and post-instruction, a large fraction ($>80\%$) of pre-service teachers stated that they believed it was possible or even very likely that life existed elsewhere in the universe, with many explicitly citing the existence of extrasolar planets (about a third pre-course, and more than two-thirds post-course) as one of the factors that bolstered their belief in this possibility.  This could be ascribed to the significant coverage exoplanets have enjoyed in recent years in popular media.
\section{Discussion and conclusions} \label{sec:discuss}

This short paper has served to introduce the NIAQ, an instrument adapted from the IAQ and administered to pre-service teachers and middle school students in Norway.

The exercise of translating from English into Norwegian and vice versa highlighted how a direct translation (with a fixation on the word level) is often neither culturally correct nor linguistically appropriate. Although the lack of a true `direct translation' introduces one extra level of interpretation into the analysis, it also strengthened the analysis by revealing how much implicit interpretation must be made when trying to infer the intended meaning from the actual words that respondents put down on paper.

The NIAQ has proven to be a rich source of data, and has provided useful and sometimes unexpected insights into issues pertaining to students learning astronomy and physics in Norway. For example, our finding about the drop in pre-service teachers' \textcolor{black}{pedagogical} behavior, with a corresponding move towards \textcolor{black}{authoritative} responses, is of particular concern given that these future teachers might na\"ively be expected to adopt pedagogically sound approaches to explaining content or concepts to their future students. Further study of this phenomenon is warranted. We also identified potentially serious issues relating to middle school students' conceptions of size and distances in the universe, with significant stratification along gender lines. Though not as prevalent, the issues related to size and distances also manifested in our sample of pre-service teachers. On a more positive note, however, we identified some early predictors of post-course success for the pre-service teachers, which could in future be used to identify students requiring special teaching intervention.

Despite the limitations of the instrument, we were able to probe three broad areas relating to student engagement with astronomy; and though the analysis of free-response writing was a labor-intensive and time-consuming exercise, it was clear that such a wealth of information could not have been obtained with a multiple-choice format alone.

A more detailed analysis of the results of the NIAQ will be the focus of future work.

\begin{acknowledgments}
The authors thank Hilde Midtg{\aa}rd Stephansen and Ina Camilla Lauvli Engan for feedback on the wording of the NIAQ, \textcolor{black}{and thank the anonymous referees for constructive comments that helped improve the paper}. V.R.\ and M.C.E.\ are grateful to the Rhodes Trust for financial support, which made this work possible; V.R.\ further acknowledges the American Association of Physics Teachers and the National Research Foundation of South Africa for financial support.
\end{acknowledgments}

\bibliography{IAQ-paper}
\end{document}